 \documentstyle[aps,eqsecnum]{revtex}

\begin{document}
\title{Monge Distance between Quantum States}

\author{Karol {\.Z}yczkowski$^1$\footnote{Fulbright Fellow. Permanent
address: Instytut Fizyki im. Mariana Smoluchowskiego,
Uniwersytet Jagiello{\'n}ski, ul. Reymonta 4, 30-059 Krak{\'o}w, Poland}
 and Wojciech S{\l}omczy{\'n}ski$^2$}
\address{
$^1$Institute for Plasma Research, University of Maryland,
College Park, MD 20742-3511  U.S.A.}
\address{
$^2$Instytut Matematyki, Uniwersytet Jagiello\'nski, 
ul. Reymonta 4, 30-059 Krak\'ow, Poland}

\date{\today}

\maketitle

\vskip 0.4cm
\begin{center}
{\small e-mail:
$^1$karol@ipr.umd.edu \quad
 $^2$slomczyn@im.uj.edu.pl }
\end{center}

\vskip 0.5cm
\begin{abstract}
We define a metric in the space of quantum states taking the Monge distance
between corresponding Husimi
distributions (Q--functions). This quantity
fulfills the axioms of a metric
 and satisfies the following {\sl semiclassical
property}: the distance between two coherent states is equal to the
Euclidean distance between corresponding points in the classical phase
space. We compute analytically distances between certain states (coherent,
squeezed, Fock and thermal) and discuss a scheme for numerical computation of
Monge distance for two arbitrary quantum states.
\end{abstract}
\vskip 1.0cm

\newpage

\section{Introduction}

The state space of an $n$-dimensional quantum system is the set of all $%
n\times n$ positive semidefinite complex matrices of trace 1 called {\sl %
density matrices}. The density matrices of rank one ({\sl pure states}) can
be identified with nonzero vectors in a complex Hilbert space of dimension $%
n $. However, one has to take into account that the same state is described
by a vector $\psi $ and $\lambda \psi $, where $\lambda \neq 0$. Hence pure
states are in one-to-one correspondence with rays $\left\{ \lambda \psi
:0\neq \lambda \in {\bf C}\right\} $. The rays form a smooth manifold called
complex projective space ${\bf CP}^{n-1}$. In the infinite-dimensional case
we have to consider density operators instead of density matrices, and the
space of pure states is the complex projective space over the
infinite-dimensional Hilbert space.

The problem of measuring a distance between two quantum states with a
suitable metric attracts a lot of attention in recent years. The
Hilbert-Schmidt norm of an operator $||A||_2=\sqrt{{\rm {Tr}(A^{\dagger }A)}}
$ induces a natural distance between two density operators $d_{HS}(\rho
_1,\rho _2)=\sqrt{{\rm {Tr}}\left[ (\rho _1-\rho _2)^2\right] }$. This
distance has been recently used to describe the dynamics of the field in
Jaynes--Cummings model \cite{ko95} and to characterize the distance between
certain states used in quantum optics \cite{wu95}. Another distance
generated by the {\sl trace norm} $||A||_1={\rm Tr}\sqrt{A^{\dagger }A}$ was
used by Hillery \cite{h87,h89} to measure the non-classical properties of
quantum states.

A concept of {\sl statistical distance} was introduced by Wootters \cite{w81}
in the context of measurements which optimally resolve neighboring pure
quantum states. This distance, leading to the Fubini-Study metric in the
complex projective space, was later generalized to density matrices by
Braustein and Caves \cite{bc94} and its dynamics for a two-state system
was studied by Braunstein and Milburn \cite{bm95}. It was shown
\cite{bc94} that for neighboring
density matrices the statistical distance is proportional to the distance
introduced by Bures in late sixties \cite{b69}. The latter was studied by
Uhlmann \cite{u76} and H{\"{u}}bner \cite{h92}, who found an explicit
formula for the Bures distance between two density operators $d_B^2(\rho
_1,\rho _2)=2(1-\rm{tr}
[(\rho _1^{1/2}\rho _2\rho _1^{1/2})^{1/2}])$.
Note that
various Riemannian metrics on the spaces of quantum states were also
considered by many other authors (see \cite{u94,d95,ps95,u95,s96}).

In the present paper we introduce yet another distance in the space of
density operators, which fulfills a following {\sl semiclassical
property}
: the distance between two coherent states
 $|\alpha_1\rangle $ and $|\alpha_2\rangle $
localized at points ${a_1}$  and $a_2$ of the classical phase space
$\Omega $
endowed with a metric $d$ is equal to the distance of these points

\begin{equation}
D(|\alpha_1\rangle ,|\alpha_2\rangle )=d (a_1,a_2).
\label{clasprop}
\end{equation}

This condition is quite natural in the semiclassical regime, where the
quasi-probability distribution of a quantum state tends to be strongly
localized in the vicinity of the corresponding classical point. A motivation
to study such a distance stems from the search for quantum Lyapunov
exponent, where a link between distances in Hilbert space and in the
classical phase space is required \cite{zsw94}.

In order to find a metric satisfying condition (\ref{clasprop}) it is
convenient to represent a quantum state $\rho $ in the Q--representation
(called also the Husimi function) \cite{h40}

\begin{equation}
H_{\rho}(\alpha ):=\langle \alpha |\rho |\alpha \rangle, \label{hus1}
\end{equation}
defined with the help of a family of coherent states $|\alpha \rangle $, 
$\alpha \in \Omega $, which fulfill the identity resolution ${\bf I}
=\int_\Omega |\alpha \rangle \langle \alpha |dm(\alpha )$, where $m$ is the
natural measure on $\Omega $. For a pure state $|\psi \rangle $ one has $
H_\psi (\alpha )=|\langle \psi |\alpha \rangle |^2$.
 The choice of coherent
states is somewhat arbitrary and in fact one may work with different systems
of coherent states \cite{z87}. In the present paper we use the classical
{\sl harmonic oscillator (field) coherent states},
 where $\Omega ={\bf C}$
and $dm(\alpha )= d^2\alpha / \pi $. For convenience we will use in the
sequel the renormalized version of the Husimi function putting $H(\alpha
)=\langle \alpha |\rho |\alpha \rangle / \pi $ and integrating over
$d^2\alpha $. Note that the Husimi representation of a given density operator
determines its uniquely \cite{t86}. Since Husimi distributions are
non-negative and normalized, i.e., $\int_\Omega H(\alpha )dm(\alpha )=1$
and $ H\geq 0$, 
it follows that a metric in the space of probability densities
$ Q:\Omega \to R_{+}$ induces a metric in the state space.

In this work we propose to measure the distance between density operators by
the Monge distance between the corresponding Husimi distributions. The
original Monge problem consists in finding an optimal way of transforming a
pile of sand into a new location. The Monge distance between two piles is
given by the `path' traveled by all
 grains under the optimal transformation
\cite{r84,r91}.

This paper is organized as follows. In sect.~II we give a definition of the
Monge metric, present an explicit solution for the one--dimensional case and
discuss some methods of tackling the two--dimensional problem. Sect.~III
contains some examples of computing the Monge distance between certain
states often encountered in quantum optics. Concluding remarks are
given in sect.~IV.
 Variational approach to the Monge problem is briefly presented in 
the Appendix.

\section{Monge metric}

\subsection{Monge transport problem}

The original Monge problem, formulated in 1781 \cite{monge},
emerged from studying the most efficient way of transporting soil
\cite{r84}:

{\sl Split two equally large volumes into infinitely small particles and
then associate them with each other so that the sum of products of these
paths of the particles to the volume is least. Along what paths must the
particles be transported and what is the smallest transportation cost?}

Fig.~1 represents a scheme for
this problem. Here we give a general definition of the Monge distance
between two probability densities $Q_1$ and $Q_2$ defined in $S=R^n$. Let 
$\Omega _1$ and $\Omega _2$, determined by $Q_i$, describe the initial and
the final location of `sand': $\Omega _i=\{\left( x,y\right) \in S\times
R^{+}:0\le y\leq Q_i(x)\}$. Due to normalization of $Q_i$ the integral
 $\int_{\Omega _i} d^nx\,dy$ \/is equal to unity. Consider
$C^1$   maps
$T:S\to S$ which generate volume preserving transformations
 $\Omega _1$ into $\Omega _2$,  i.e.,
\begin{equation}
Q_1\left( x\right) =Q_2\left( Tx\right) \left| T'(x)\right|
\label{trans}
\end{equation}
for all $x\in S$, where $T'(x)$ denotes the Jacobian of the
map $T$ at point $x$.
We shall look for a transformation giving the
minimal
displacement integral and define the Monge distance \cite{r84,r91}
\begin{equation}
D_M(Q_1,Q_2):={\rm inf}\int_S|x-T(x)|Q_1\left( x\right) d^nx\ \text{,}
\label{monge}
\end{equation}
where the infimum is taken over all
 $T$ as above. The optimal transformation
(if exists) $T_M$ is called a {\sl Monge plan}. Note that in this
formulation of the problem the ''vertical'' component of the sand movement
is neglected.

It is easy to see that Monge distance fulfills all the axioms of a metric.
This allows us to define a {\sl 'classical'} distance between two
quantum states $\rho_1$ and $\rho_2$ via the Monge distance between the
corresponding Husimi distributions:
\begin{equation}
D_{cl}(\rho_1,\rho_2):= D_M\bigl(H_1(\alpha), H_2(\alpha) \bigr).
\end{equation}

The Monge distance satisfies the semiclassical property: it
is shown below that the
distance between two coherent states, represented by Gaussian Husimi
distributions localized at points $a_1$ and $a_2$, equals to the classical
distance $|a_1-a_2|$.

It is sometimes useful to generalize the notion of the Monge metric and to
define a family of distances $D_{M_p}$ labeled by a continuous index $%
p~\left( 0<p\le \infty \right)$ in an analogous way:
\begin{equation}
\lbrack D_{M_p}(Q_1,Q_2)]^p:={\rm inf} \int_S|x-T(x)|^pQ_1\left( x\right)
d^nx\
\text{.}  \label{mongep}
\end{equation}
For $p=1$ one recovers the Monge distance $D_{M_1}=D_M$,
while the Fr{\'e}chet distance $D_{M_2}$ is obtained for $p=2$.
A more general approach to the
transport problem was proposed by Kantorovitch \cite{k39}
 and further developed by Sudakov \cite{sud}.
In contrast with the definition of Monge discussed in this work,
the Kantorovitch distance between $Q_1$ and $Q_2$ is explicitly
symmetric with respect to exchange of both distributions.
For a comprehensive review of metrics in the space
of probability measures and other generalizations
 of the Monge distance see
the monograph of Rachev \cite{r91}.

\subsection{Salvemini solution for 1D problem}

For $S=R$ the Monge distance can be expressed explicitly with the help of
distribution functions $F_i(x)=\int_{-\infty }^xQ_i(t)dt$, $i=1,2$.
Salvemini \cite{s43} and Dall'Aglio \cite{a56} obtained the following
solution of the problem for $p=1$:
\begin{equation}
D_M(Q_1,Q_2)=\int_{-\infty }^{+\infty }|F_1(x)-F_2(x)|dx\text{ ,}
\label{salv}
\end{equation}
represented schematically in Fig. 2.

This result was generalized in  the
fifties to all $p\geq 1$ by several authors (see
\cite{r84,r91}). We have

\begin{equation}
\lbrack D_{Mp}(Q_1,Q_2)]^p=\int_0^1|F_1^{-1}(t)-F_2^{-1}(t)|^pdt\text{ . }
\label{salv2}
\end{equation}

\subsection{Poisson-Ampere-Monge equation for 2D case}

Consider smooth densities $Q_1$,
$Q_2:R^2\rightarrow R^{+}$. We are looking
for a transformation field $w=\left( w_1,w_2\right) :R^2\rightarrow R^2$
fulfilling $w(x_1,x_2)=T_M(x_1,x_2)-(x_1,x_2)$, where $T_M$ is an optimal
Monge
transformation minimizing the right-hand side of (\ref{mongep}). Restricting
ourselves to smooth transformations $T$ we may apply the standard
variational search for the optimal field $w$. In Appendix it is shown that
${\rm rot}(w)=0$ if $p=2$, and the potential $\varphi :w={\rm grad}(\varphi )
$ satisfies the following partial differential equation

\begin{equation}
\varphi
_{x_1x_1}+\varphi_{x_2x_2}+\varphi_{x_1x_1}\varphi_{x_2x_2}-(\varphi
_{x_1x_2})^2={%
\frac{Q_1(x_1,x_2)}{Q_2(x_1+\varphi_{x_1},x_2+\varphi_{x_2})}}-1\text{ .}
\label{LAM}
\end{equation}
Solving this {\sl Laplace--Ampere--Monge (LAM) equation} for the potential $%
\varphi $ we get the Fr{\'e}chet  distance $D_{M_2}$ from (\ref{mongep})
computing the minimal displacement
\begin{equation}
(D_{M_2}(Q_1,Q_2))^2=\int_{-\infty }^{+\infty }dx_1\int_{-\infty
}^{+\infty
}dx_2\,(\varphi_{x_1}^2+\varphi_{x_2}^2)Q_1(x_1,x_2)\text{ .}
\label{monge2}
\end{equation}
Though it is hardly possible to solve equation (\ref{LAM}) in a general
case, it might be used to check, whether a given transformation can be a
solution of the Monge problem.
Let us remark that an optimal transformation field $w$ (if exists) must not
be unique. Moreover, (\ref{LAM}) provides only a sufficient condition for $w$
being optimal.
It is important to note that the two dimensional Monge problem posed two
hundreds years ago has not been solved in a general case until now \cite
{r84,r91}.

\subsection{Estimation of the Monge distance via transport problem}

One of major tasks of linear programming is the optimization of the
following transport problem. Consider $N$ suppliers producing $a_i$ $\left(
i=1,\dots ,N\right) $ pieces of a product per a time period and $M$
customers requiring $b_j$ $\left( j=1,\dots ,M\right) $ pieces of the
product at the same time. Let $\left( c_{ij}\right) $ be a $N\times M$ cost
matrix, representing for example the distances between sites. Find the
optimal transporting scheme, minimizing the total transport costs $%
C=\sum_{i=1}^N\sum_{j=1}^Mc_{ij}x_{ij}$. The non-negative elements of
the unknown matrix $\left( x_{ij}\right) $ denote
 the number of products moved
from $i$-th supplier to $j$-th customer.
The optimization problem is subjected
to the following constrains: $\sum_{j=1}^Mx_{ij}=a_i$ and
$\sum_{i=1}^Nx_{ij}=b_j$.
In the simplest case the total supply equals the total demand and $%
\sum_{i=1}^Na_i=\sum_{j=1}^Mb_j$.

The {\sl transport problem} described above gave, with a related {\sl
assignment problem}, an impulse to develop methods of linear programming
already half a century ago \cite{k39}, \cite{h41}. Since then several
algorithms for solving the transport problem numerically have been proposed. Some
of them, as the {\sl northwest corner} {\sl procedure} and {\sl Vogel's
approximation} \cite{wc91}, are available in specialized software
packages.
It is worth to add that the transport methods are widely used to solve a
variety of problems in business and economy such as, for instance, market
distribution, production planning, plant location and scheduling problems.

It is easy to see that the transport problem is a discretized version of the
Monge problem defined for continuous distribution functions. One can,
therefore, approximate the Monge distance between two distributions
$Q_1(x)$ and $Q_2(x)$ (where $x$ stands for a two dimensional vector),
 by solving the transport problem for delta peaks
approximation of 
the continuous distributions: $q_1=\sum_{i=1}^NQ_1(x_i)%
\delta (x-x_i)$ and $q_2=\sum_{j=1}^MQ_2(x_j)\delta
(x-x_j)$.
The quality of this approximation depends on the numbers $N$ and $M$
of peaks
representing the initial and the final distribution,
 respectively, and also on
their location with respect to the shape of both distributions. Numerical
study performed with the {\sl northwest corner} algorithm for some
analytically soluble examples of the 2-D Monge problem shows \cite{swz97}
that for reasonably smooth distributions one obtains Monge distance with
fair accuracy for a number of peaks of the order of hundreds.

\section{Monge distance between some states of quantum optics}

In this section we compute Monge distances between certain quantum states.
Even though our considerations are valid in the general framework of
quantum
mechanics, for the sake of concreteness we will use the language of quantum
optics. Let us recall briefly the necessary definitions and properties.

Let $a$ and $a^{\dagger }$ be the annihilation and creation operators
satisfying the commutation relation $[a,a^{\dagger }]=1$. The
{\sl 'vacuum'} state
$|0\rangle $ is distinguished by the relation $a|0\rangle =0$. Standard
harmonic-oscillator {\sl coherent states}
$|\alpha \rangle $ can be defined as the 
eigenstates of the annihilation operator $a|\alpha \rangle =\alpha |\alpha
\rangle $ or by the Glauber translation operator $D(\alpha )=\exp (\alpha
a^{\dagger }-\alpha ^{*}a)$ as $|\alpha \rangle =D(\alpha )|0\rangle $.
 Coherent states, determined by an arbitrary complex
number $\alpha =x_1+ix_2$, enjoy a minimum uncertainty property. They are
not orthogonal and do overlap. The Husimi distribution of a
coherent state $|\beta \rangle $ is Gaussian
\begin{equation}
H_\beta (\alpha )={\frac 1\pi }|\langle \beta |\alpha \rangle |^2=
{\frac 1\pi }\exp (-|\alpha -\beta |^2).
 \label{huscoh}
\end{equation}
{\sl Squeezed states}
 $|\gamma ,\alpha \rangle $ also minimize the uncertainty
relation, however the variances of both canonically coupled variables are
not equal. They are defined as
\begin{equation}
|\gamma ,\alpha \rangle :=D(\alpha) S(\gamma )|0 \rangle ,
\label{sq1}
\end{equation}
where the squeezing operator is $S(\gamma )=
{\rm \exp }[{1\over 2}(\gamma^* a^2- \gamma a^{\dagger 2})]$.
 The modulus $g$ of the
complex number $
\gamma =ge^{2i\theta }$ determines the
strength of squeezing, $s=e^g
-1$, while the angle
$\theta $ orients the squeezing axis.
 The Husimi distribution of a
squeezed
state $|\gamma ,\beta \rangle $ is a non-symmetric Gaussian localized at
point $\beta $ and for $\theta=0$ reads
\begin{equation}
H_{\gamma ,\beta }(x_1,x_2)={\frac 1\pi } \exp
 [-({\rm{Re}}(\beta)-x_1)^2 /(s+1)^2-({\rm{Im}}(\beta
)-x_2)^2 (s+1)^2]  \label{hussq}
\end{equation}

Each pure state can be expressed in the Fock basis consisting of $n$--photon
states $|n\rangle $, $n=0,1,2,\dots $. Each coherent state can be expanded
in the Fock basis as
\begin{equation}
|\alpha \rangle =e^{-|\alpha |^2/2}\sum_{n=0}^\infty {\frac{\alpha ^n}
{\sqrt{ n!}}}|n\rangle .  \label{cohexp}
\end{equation}
The known scalar product $\langle \alpha |n\rangle $ allows one to
write the Husimi function of a {\sl Fock state}
\begin{equation}
H_{|n\rangle }(\alpha )={\frac 1\pi }{\frac{|\alpha
|^{2n}}{n!}}e^{-|\alpha|^2}.  \label{fockexp}
\end{equation}
The {\sl vacuum state} $|0\rangle $ can be thus regarded as the single
Fock state being simultaneously coherent.

In contrast to the above mentioned pure states, the {\sl thermal}
mixture of states  with the
mean number of photon equal to $\bar{n}$ is represented by the
density operator
\begin{equation}
\rho_{\bar n}=\sum_{n=0}^{\infty} {\frac{{\bar n}^n }{( {\bar n}+1)^{n+1}}}
|n\rangle \langle n|.  \label{term1}
\end{equation}
Its Husimi distribution is the Gaussian centered at zero with the width
depending on the mean photon number,

\begin{equation}
H_{\rho _{\bar{n}}}(\alpha )={\frac 1{\pi ({\bar{n}}+1)}}\exp \left(
-{\frac{ |\alpha |^2}{{\bar{n}}+1}}\right) .  \label{termhus}
\end{equation}

\subsection{Two coherent states}

Let us consider an arbitrary two dimensional distribution function
$Q_1(x_1,x_2)$
and a shifted one $Q_2(x_1,x_2)=Q_1(x_1-x_1^*,x_2-x_2^*)$. It is
intuitive to expect
that the simple translation given by $w(x_1,x_2)={\rm
{const}}=(x_1^*,x_2^*)$ solves 
the corresponding Monge problem. Since for the respective potential $\phi
(x_1,x_2)=x_1^* x_1+x_2^*x_2$ the second derivatives vanish, the both
two sides of
the LAM equation (\ref{LAM}) are equal to zero and the maximization 
condition is fulfilled.

It follows from this observation that for two coherent states defined on the
complex plane the Monge plan reduces to translation. Integration in (\ref
{monge}) is trivial and provides the Monge distance between two arbitrary
coherent states $|\alpha \rangle $ and $|\beta \rangle $
\begin{equation}
D_{cl}(|\alpha \rangle ,|\beta \rangle )=|\alpha -\beta |.  \label{discoh}
\end{equation}
This is exactly the{\sl \ semiclassical property} we demanded from the
metric in the state space. The distance of the coherent state $|\alpha
\rangle $ from the vacuum state $|0\rangle $ is equal to $|\alpha |$,
which simply is the square root of the mean number of photons
in the state $|\alpha \rangle $. The classical property is fulfilled by
the generalized Monge metric $D_{M_p}$ for any positive parameter $p$.

\subsection{Coherent and squeezed states}

We compute the distance between a coherent state $|\alpha \rangle $ and the
corresponding state squeezed $|\gamma ,\alpha \rangle $. Due to invariance
of the Monge optimization with respect to translations this distance is
equal
to the distance between vacuum $|0\rangle $ and the squeezed vacuum $|\gamma
,0\rangle $. For simplicity we will assume that squeezing is performed along
the $x_1$ axis, i.e., the complex squeezing parameter is real $\gamma
=g\in R$.

The corresponding Monge problem consists in finding the optimal
transformation of the symmetric
 Gaussian $Q_1(x_1,x_2)=\exp (-x_1^2-x_2^2)/\pi $
into an asymmetric one $Q_2(x_1,x_2)=\exp (-x_1^2/ (s+1)^2 -x_2^2
(s+1)^2)/\pi$.
Considering contours of the Husimi distribution, often used to
represent a state in quantum optics, a
 circle has to be transformed into an
ellipse. If $p=2$ then the following affine transformation $
T(x_1,x_2)=(x_1/(s+1),x_2(s+1))$ corresponds to
 the irrotational transport field 
$w(x_1,x_2)=(-sx_1/(s+1),sx_2)$. It can be obtained as the gradient of
the potential 
$\varphi (x_1,x_2)=-sx_1^2/(2s+2)+sx_2^2/2$, for which both sides of the
LAM
equation (\ref{LAM}) vanish. Hence field ${w}$ provides a Monge plan for
this problem and the distance is given by its length $|{w|}$ integrated over
the volume of $Q_1$

\begin{equation}
D_{cl}(|0\rangle ,|g,0\rangle )={\frac s\pi }\int_{-\infty }^\infty
dx_1 \int_{-\infty }^\infty dx_2\,\exp
(-x_1^2-x_2^2)\sqrt{x_1^2/(s+1)^2+x_2^2}
={\frac s{ \sqrt{\pi }}E}
\left( \frac{s^2+2s}{s^2+2s+1}\right) \text{{\rm \ ,}}
\label{dissq}
\end{equation}
where $E(x)$ stands for the complete elliptic integral of the second kind.

A simpler result may be obtained in this case for the Fr{\'e}chet distance 
$D_{M_2}$

\begin{equation}
\bigl[D_{M_2}(|0\rangle ,|g,0\rangle )\bigr]^2 = {\frac {s^2}\pi
}\int_{-\infty
}^\infty dx_1 \int_{-\infty }^\infty dx_2\exp (-x_1^2-x_2^2)
(x_1^2/(s+1)^2+x_2^2)={\frac{s^2}2}
(1+{\frac 1{(s+1)^2}}).  \label{dissq2}
\end{equation}
Eventually, the `Chebyshev' distance $D_{M_\infty }$, characterizing the
maximal dislocation length, is equal to $s$. For large squeezing all three
distances grow proportionally to the  squeezing parameter $s$,
 with slopes 
$k_1=1/\sqrt{\pi }$, $k_2=1/\sqrt{2}$ and $k_\infty =1$ ordered according
to the index $p$. This agrees with an observation that $p_1<p_2$ implies
$ D_{M_{p_1}}\le D_{M_{p_2}}$ \cite{r91}.

\subsection{Vacuum and thermal states}

Since Husimi distributions of both states is rotationally invariant, it is
convenient to use radial components of the distribution $R_i(r)=2\pi
rH_i(r,\phi )$. One may then reduce the problem to one dimension and
find the Monge distance via radial distribution functions 
$F_i(r)=\int_0^rR_i(r^{\prime })dr^{\prime }$. Taking Husimi distributions
(\ref{huscoh}) and (\ref{termhus}) we get the corresponding distribution
functions $F_1(r)=1-\exp (-r^2)$ and $F_2(r)=1-\exp \bigl(-r^2/({\bar{n}}+1)
\bigr)$. Using the Salvemini formula (\ref{salv}) we obtain
\begin{equation}
D_{cl}(|0\rangle \langle 0|,\rho _{\bar{n}})=\int_0^\infty
|F_1(r)-F_2(r)|dr={%
\frac{\sqrt{\pi }}2}\bigl( \sqrt{{\bar{n}}+1}-1\bigr).
\end{equation}
In the same way we get a more general formula for the Monge distance between
two thermal states
\begin{equation}
D_{cl}(\rho _{\bar{n}_1},\rho _{\bar{n}_2})={\frac{\sqrt{\pi }}2}\bigl|
\sqrt{{ \bar{n}_1}+1}-\sqrt{{\bar{n}_2}+1}\bigr|.
\end{equation}

\subsection{Two Fock states}

As in the previous example the rotational symmetry of the Fock states allows
us to use the 1D formula. Integrating (\ref{fockexp})
for a Fock state $|n\rangle $ we can express
the radial distribution function in terms of the incomplete Gamma
function $\Gamma(x,r)$ as
\begin{equation}
F_n(r)=1-{\frac{\Gamma (n+1,r^2)}{\Gamma (n+1)}}.  \label{fo1}
\end{equation}
Since for different Fock states the distribution functions do not cross,
applying Salvemini formula (\ref{salv}) we obtain the Monge
distance
\begin{equation}
D_{cl}(|m\rangle ,|n\rangle )=|\int_0^\infty F_n(r)dr-\int_0^\infty
F_m(r)dr|
= \sqrt{\pi}|C_m-C_n|,
\end{equation}
where the integrals $C_i$ can be found analytically:
$C_0=1/2,~C_1=3/4,~C_2=15/16,~ C_3=35/32,...$.

\section{Discussion}

We have presented a definition of distance between quantum states (i.e.
elements of a Hilbert space) which possesses a certain  classical
property, natural
for investigation of the semiclassical limit of quantum
mechanics. Monge
distance between the corresponding
 Husimi functions fulfills the axioms of a metric
and induces a 'classical' topology in the Hilbert space.
It is worth to emphasize that the Monge distance between
 two given density matrices depends on the
 topology of the corresponding classical phase space.

Consider a quantum state prepared as a superposition of two coherent
states  separated in the phase space by $x$. It is known
\cite{z86,h91,zp94} that such a state
interacting with an environment
evolves quickly toward a mixture of the two localized states.
The decoherence time decreases with the classical distance $x$, equal
just to the Monge distance between both coherent states.
We expect therefore that the Monge distance
between two arbitrary quantum states might be useful
to determine
the rate with which the superposition of these two states suffers the
decoherence.

It is possible to generalize
our approach in several directions. Instead of the
standard Husimi distributions used throughout this paper, one may study
Monge distances between generalized Husimi distributions 
$\tilde{H}_{\rho}(\alpha)=\langle 
\tilde{\alpha}|\rho |\tilde{\alpha}\rangle, $ where
$|\tilde{\alpha} \rangle $
 are generalized coherent states \cite{p86}. For example, one may
use for this purpose squeezed states \cite{wh87}, or the spin coherent
states \cite{r71,a72}, if the classical space
is the two-dimensional sphere.

Moreover, our considerations based on the Husimi representation of quantum
states, may be in fact extended to the Wigner function. In spite of the fact
that the Wigner function can take on negative values, it is normalized and
the Monge problem of finding an optimal way to transport one Wigner function
into another might also be considered. The concept of the
classical distance between two
Husimi (Wigner) functions is not only of theoretical interest, since novel
methods of measuring Husimi and Wigner distributions have been recently 
developed \cite{vr89,sbrf93,a95}.

\section{Acknowledgments}

We thank Harald Wiedemann for fruitful cooperation and 
Fritz Haake for valuable remarks. One of us (K.{\.Z}) is
grateful to Isaac Newton Institute for Mathematical Sciences in
Cambridge, where part of this work was done.  Financial
support by the Polish KBN grant P03B 060 13 is gratefully acknowledged.

\appendix

\section{Variational approach to Monge problem}

\subsection{1D case}

Let $Q_1$ and $Q_2$ be smooth densities. In 1D case there is only one map 
$T$ fulfilling (\ref{trans}). It can be described with the aid of
distribution functions as $T(x)=F_2^{-1}(F_1(x))$, where 
$F_i(x)=\int_{-\infty }^xQ_i(t)dt $ for $x\in R$.
This allows us to express the generalized Monge distance as an integral (\ref
{mongep})
\begin{equation}
\lbrack D_{Mp}(Q_1,Q_2)]^p:=\int_{-\infty }^\infty
Q_1(x)|F_2^{-1}[F_1(x)]-x|^pdx.  \label{monp22}
\end{equation}
In the simplest case $p=1$ (\ref{monp22}) reduces to the Salvemini formula (%
\ref{salv}) and for $p>1$ to formula (\ref{salv2}).

\subsection{2D case}

Consider two smooth densities $Q_1(x_1,x_2)$ and $Q_2(x_1,x_2)$. We are
looking for a map 
$T(x_1,x_2)=(x+w_1(x_1,x_2),x_2+w_2(x_1,x_2))$ transforming $Q_1$ into
$Q_2$ (i.e. such
that (\ref{trans}) is fulfilled) and minimizing the quantity
\begin{equation}
I_p=\int_{-\infty }^{+\infty
}Q_1(x_1,x_2)|w_1^2(x_1,x_2)+w_2^2(x_1,x_2)|^{p/2}dx_1 dx_2
\text{ .}
\end{equation}
The index $p$, labeling the generalized distance, is equal to one for the
Monge metric and to two for the Fr{\'e}chet metric.
Introducing a Lagrange factor $\lambda $ we write the Lagrange
function in the form
\begin{equation}
L_p=Q_1(w_1^2+w_2^2)^{p/2}+\lambda
(Q_1-Q_2(T))\bigl[(1+w_{1x_1})(1+w_{2x_2})-w_{1x_2}w_{2x_1}\bigr] .
\end{equation}
The Lagrange--Euler equations for variations of $L_p$ allow us to obtain the
partial derivatives of $\lambda $
\begin{eqnarray}
\lambda _{x_1} &=&2pC_p(w_1(1+w_{1x_1})+w_2w_{2x_1})\text{ ,}  \nonumber
\\
\lambda _{x_2} &=&2pC_p(w_2(1+w_{2x_2})+w_1w_{1x_2})\text{ ,}
\label{lambda}
\end{eqnarray}
where $C_p= (w_1^2+w_2^2)^{(p-2)/2}$. Using the equality
$\lambda_{x_1x_2}=\lambda _{x_2x_1}$ we get

\begin{equation}
w_{1x_2}\left( w_1^2(p-1)+w_2^2\right) -w_{2x_1}\left(
w_2^2(p-1)+w_1^2\right)
+(p-2)(w_{2x_2}-w_{1x_1})w_1w_2=0\text{ .}  \label{cond}
\end{equation}

If $p=2$ we deduce from (\ref{cond}) that $w_{1x_2}=w_{2x_1}$, i.e.,
rot$(w)=0$.
Taking the potential $\varphi :w=\rm{grad}(\varphi )$ we see that $\varphi$
fulfills (\ref{LAM}) and formula (\ref{monge2}) holds.



\begin{figure}[tbp]
\caption{ Monge transport problem: How to shovel a pile of sand
$Q_1(x_1,x_2)$ into a new location $Q_2(x_1,x_2)$ minimizing the work
done?}
\label{f1}
\end{figure}

\begin{figure}[tbp]
\caption{ Monge distance between 1D functions $Q_1(x)$ and $Q_2(x)$ may be
represented as the area between graphs of the 
corresponding distribution functions
$F_1(x)$ and $ F_2(x)$}
\label{f2}
\end{figure}

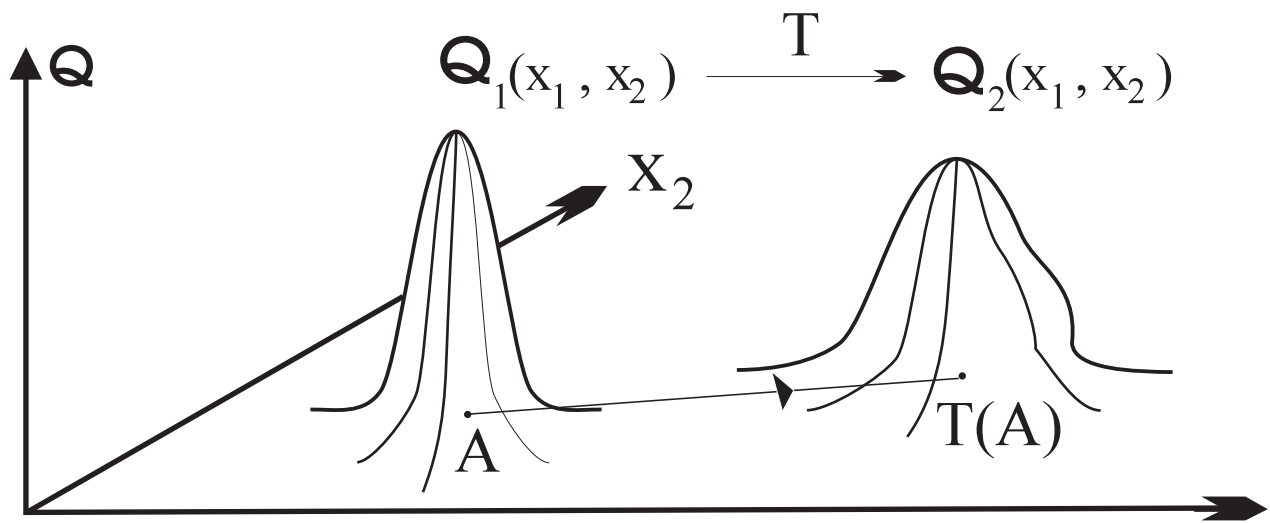
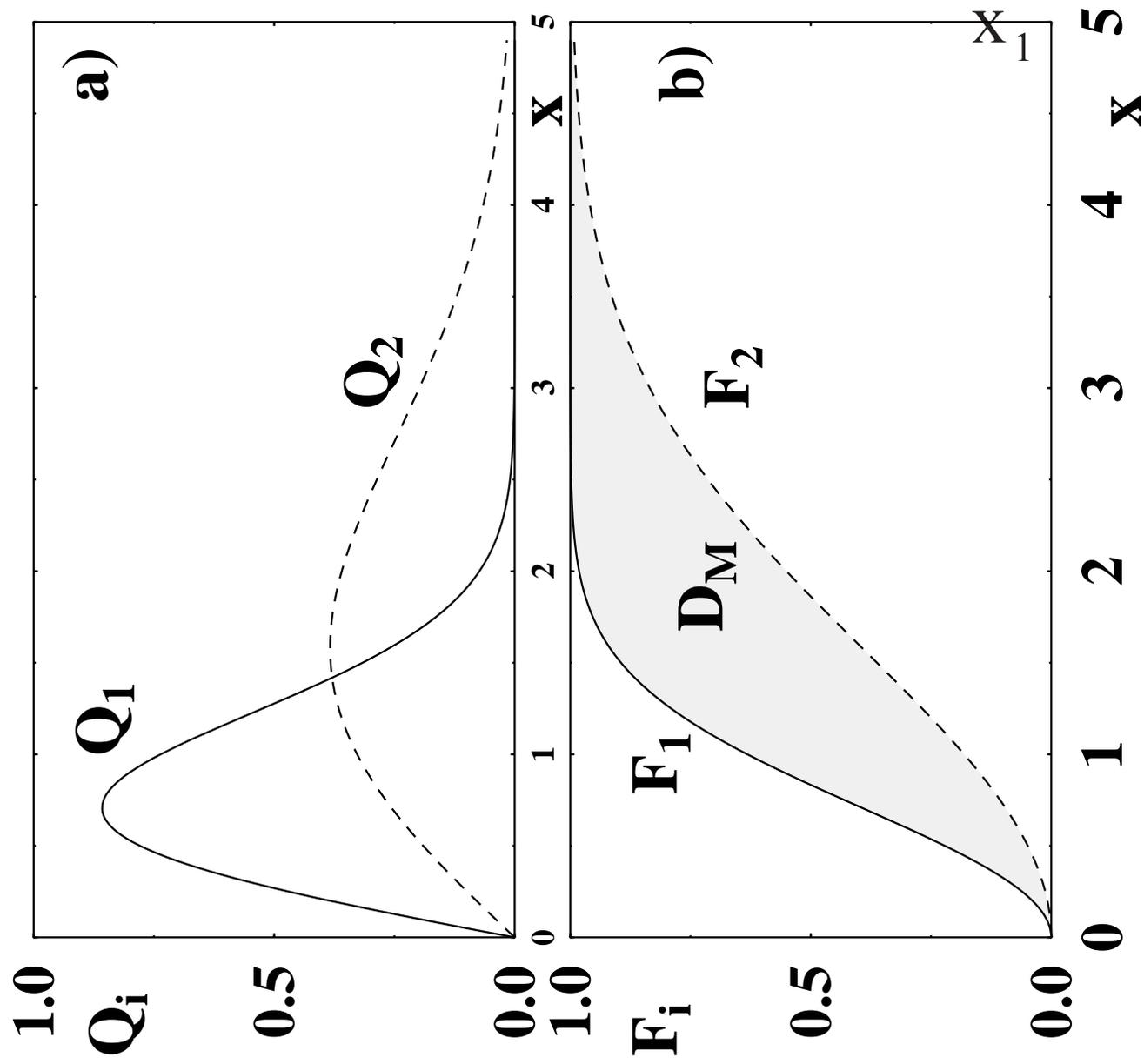
\end{document}